\documentclass[aps,twocolumn,nofootinbib,prl]{revtex4-1}

\usepackage{bm, amsmath,amssymb,amsfonts,slashed,graphicx,array,soul}
\usepackage[vcentermath,noautoscale]{youngtab}
\usepackage{bbold}
\usepackage{mathrsfs}

\newcommand{\Tr}{\mbox{Tr}}

\newcommand{\LambdaH}{\Lambda_{\textrm{H}}}
\newcommand{\LambdaF}{\Lambda_{\textrm{F}}}
\newcommand{\pda}{{\phantom{\dagger}}}

\usepackage{hyperref}	
\allowdisplaybreaks[1]

\begin{document}

\title{Disentangling Mass and Mixing Hierarchies}

\author{Simon Knapen}
\affiliation{Department of Physics, University of California, Berkeley, CA 94720, USA}
\affiliation{Ernest Orlando Lawrence Berkeley National Laboratory,
University of California, Berkeley, CA 94720, USA}
\author{Dean J. Robinson}
\affiliation{Department of Physics, University of California, Berkeley, CA 94720, USA}
\affiliation{Ernest Orlando Lawrence Berkeley National Laboratory,
University of California, Berkeley, CA 94720, USA}

\begin{abstract}
We present a fully perturbative mechanism that naturally generates mass hierarchies for the Standard Model (SM) fermions in a flavor-blind sector. The dynamics generating the mass hierarchies can therefore be independent from the source of flavor violation, and hence this dynamics may operate at a much lower scale. This mechanism works by dynamically enforcing simultaneous diagonalization -- alignment -- among a set of flavor-breaking spurions, as well as generating highly singular spectra for them. It also has general applications in model building beyond the SM, wherever alignment between exotic and SM sources of flavor violation is desired.
\end{abstract}
\maketitle

\textbf{Introduction.}
The origin of the large mass and mixing hierarchies among the Standard Model (SM) fermions  -- the flavor puzzle -- is a significant open problem in particle physics. Attempts to resolve this problem have taken a variety of approaches. The most well-known is perhaps the Froggatt-Nielsen mechanism \cite{Froggatt:1978nt}, which assigns different charges of a pseudo-anomalous symmetry among the SM generations. It thereby can physically distinguish fermion flavors and generate a hierarchy of masses and relative mixings for them. There exist multiple alternate formulations or extensions of this general idea, that assign various types of horizontal dynamics to the SM generations (see, e.g., \cite{Fritzsch:1978wi,Georgi:1983mq,Cheng:1987rs,Kaplan:1993ej,ArkaniHamed:1999dc,Nelson:2000sn,Fritzsch:2000mf,Hall:2001zb} among many others).

Approaches of this style intrinsically link the origin of the mass and mixing hierarchies. This can lead to flavor model-building challenges. For instance, considering the first two quark generations, the Cabibbo-Kobayashi-Maskawa (CKM) quark mixing matrix element $|V_{cd}| \sim 0.2$, while the mass hierarchy $m_{u,d}/m_c \sim 10^{-3}$. Similarly in the lepton sector, the charged leptons exhibit a large mass hierarchy, while the Pontecorvo-Maki-Nakagawa-Sakata (PMNS) mixing matrix elements are all $\mathcal{O}(1)$.

In this Letter we present a mechanism that dynamically and naturally generates SM mass hierarchies without charging the SM fermions under any Froggatt-Nielsen style horizontal symmetries. The SM fermions need only be charged under their $U(3)$ flavor symmetries, and couple universally to the physics that generates their mass hierarchies. This means that the scale at which the mass hierarchies are generated, $\LambdaH$, can be independent from the scale of flavor breaking, which could have interesting phenomenological consequences. For example: $\LambdaH$ may be low enough to be detectable at the LHC; if $\LambdaH$ is nearby the electroweak scale,  the Jarlskog invariant can be large during the era of sphaleron transitions, opening up a new avenue for significant electroweak baryogenesis.

\textbf{Strategy.} Specifically, we show how to dynamically generate vacuum expectation values -- spurions -- for a set of bifundamental flavon fields, $\{\lambda_\alpha\}$, with the `\emph{aligned, spectrally disjoint and rank-1}'  pattern
\begin{align}
	\label{eqn:ASDS}
	 \langle\lambda_1\rangle = U\, \mbox{diag} \Big\{ r_1,\, 0,\, \ldots \Big\}\, V^\dagger\,,\notag\\
	 \langle\lambda_2\rangle = U\, \mbox{diag} \Big\{ 0,\, r_2,\, \ldots \Big\}\, V^\dagger\,,
\end{align}
and so on, with $U$ and $V$ unitary matrices; the same for each flavon. (This is inherently different to a rank-1 projector approach. See, e.g., \cite{Barr:1981wv}.) Applied to the SM with three generations, these spurions each break $U(3)\times U(3)$-type flavor symmetry to a different subgroup, such that collectively the flavor symmetry is broken down to baryon or lepton number. With spurions of the form in eq.~\eqref{eqn:ASDS}, one may then naturally construct mass hierarchies among the SM fermions by assigning extra symmetries or dynamical effects \emph{horizontally among the flavons}. For instance, for a set of three up-type flavons $\{\lambda_{t,c,u}\}$, bifundamental under $U(3)_Q\times U(3)_U$, the up-type SM Yukawa terms could be generated from the irrelevant operator
\begin{equation}
	\label{eqn:SMY}
	H^\dagger \bar{Q}_L \Bigg\{\! \frac{s_t}{\LambdaH} \frac{\lambda_t}{\LambdaF} +\frac{s_c}{\LambdaH} \frac{\lambda_c}{\LambdaF}+ \frac{s_u}{\LambdaH} \frac{\lambda_u}{\LambdaF} \!\Bigg\}U_R\,,
\end{equation}
where $\Lambda_F\sim \langle \lambda_\alpha\rangle$ is the scale of flavor breaking. The $s_\alpha$ are $U(3)\times U(3)$ singlet operators -- in this sense they are \emph{flavor-blind} -- that encode a hierarchy $\langle s_t \rangle \gg \langle s_c \rangle \gg \langle s_u \rangle$, generated at the scale $\LambdaH$. The up-type quark mass hierarchies follow immediately from the pattern \eqref{eqn:ASDS}, independently of the structure of the matrices, $U$ and $V$, that encode flavor-violating effects. (We focus here on up-type quarks, but the generalization to the down-type quarks and leptons follows analogously.)

This approach contrasts with Froggatt-Nielsen style horizontal charges, that are assigned directly to the SM fermions. Instead, the SM fermions are coupled universally to the flavor-blind, hierarchy-generating operators $s_\alpha$. While $\langle s_\alpha \rangle/\langle s_\beta \rangle$ is fixed by the observed SM mass hierarchies, and while the flavor-breaking scale, $\LambdaF$, is bounded below by precision flavor constraints, the hierarchy scale $\LambdaH$ is unconstrained by these effects, and could be quite low.

\begin{figure}[t]
	\includegraphics[width = 4cm]{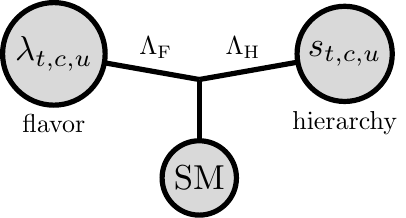}
	\hspace{1cm}
	\includegraphics[width = 3cm]{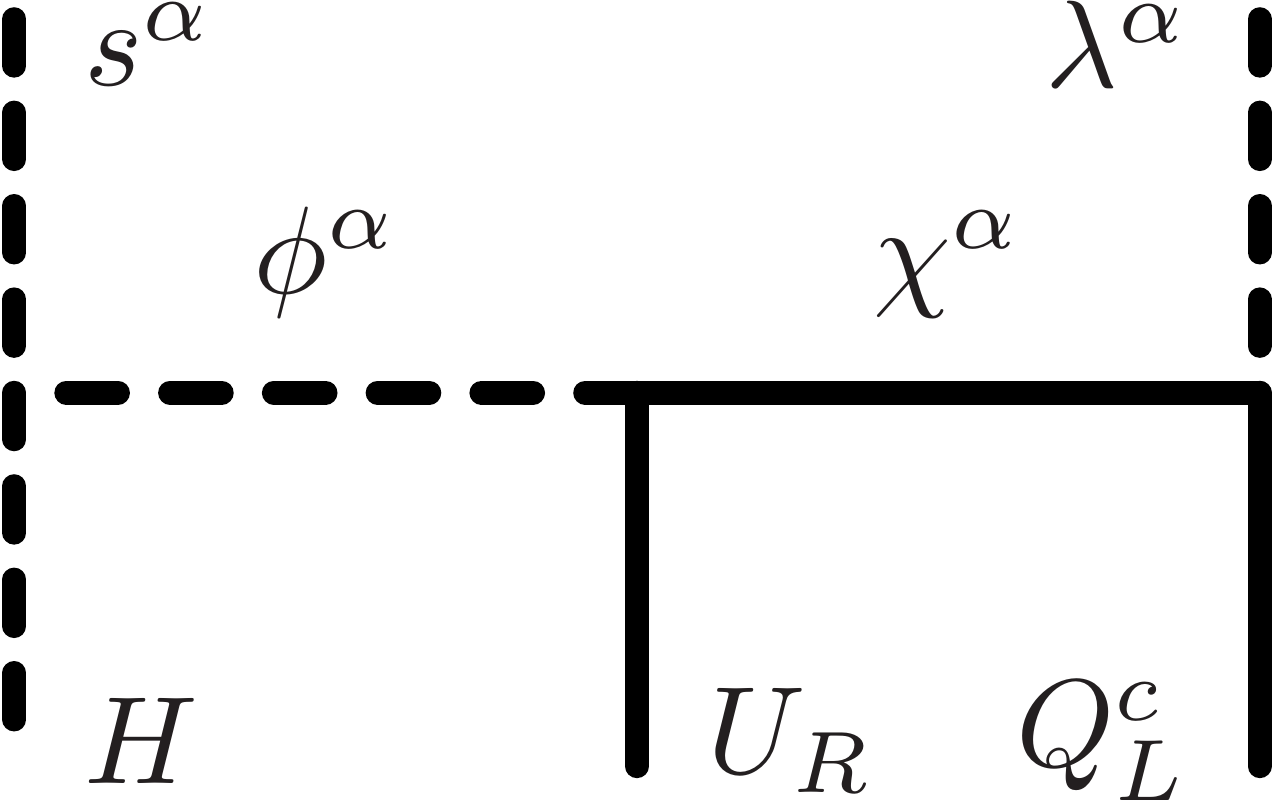}
	\caption{Left: Schematic representation of the low energy effective theory. The lines represent the irrelevant interactions in \eqref{eqn:SMY}.   Right: Sample UV completion, in which $\LambdaH$ and $\LambdaF$ are identified with the mass scales of $\phi_\alpha$ and $\chi_\alpha$ respectively. }
	\label{fig:GQ}
\end{figure}

The particular scenario we have in mind is to consider three sectors: a SM sector, a `flavor' sector, and a `hierarchy' sector. The dynamics of the flavor sector breaks the $U(3)_Q\times U(3)_U$ flavor symmetries at a scale $\LambdaF$, by generating spurions of the form \eqref{eqn:ASDS} for the three flavons $\lambda_{t,c,u}$. Suppose that these flavons also carry parity symmetries $P_\alpha: \lambda_\alpha \to -\lambda_\alpha$.  These are broken by $P_\alpha$-odd, flavor singlet spurions $\langle s_\alpha \rangle$, generated in the hierarchy sector at $\LambdaH$. The SM, flavor and hierarchy sectors then interact through the three-way portal in \eqref{eqn:SMY}. We show a schematic representation of this scenario in Fig.~\ref{fig:GQ}. We also show a sample UV completion, that generates the operator \eqref{eqn:SMY} at tree level, in which $s_\alpha$ are a set of scalar fields. (Note that other operators like $H^\dagger \bar{Q}_L{s_\alpha} \lambda_\alpha \lambda_\beta^\dagger \lambda^\pda_\beta U_R$ are annihilated by the pattern \eqref{eqn:ASDS}.) 

In the remainder of this Letter we will first specify the algebraic conditions that automatically ensure that a set of matrices is aligned, spectrally disjoint and rank-1. We proceed to show that the most general renormalizable potential for the $\lambda_\alpha$ has a minimum that enforces these algebraic conditions. With this mechanism in hand, we will present an example of a set of horizontal discrete symmetries on the spurions that can generate the SM mass hierarchies. We will also show how to construct an approximate CKM matrix within this framework.

\textbf{Algebraic Conditions.} Consider two tensors $\lambda_{iI}$ and $\xi_{iI}$ charged under the bifundamental of an $U(n)\times U(n)$ flavor symmetry group. That is $i,I = 1,\ldots,n$ are indices of the (anti)fundamental representations. We adopt a matrix notation that encodes contractions of indices of the same type. For instance $(\lambda \xi^\dagger)_{ij} \equiv \sum_I\lambda^{\phantom{*}}_{iI} \xi_{jI}^*$ and $(\lambda \xi^\dagger)^\dagger_{ij} \equiv (\xi \lambda^\dagger)_{ij}$, and similarly for uppercase indices. Hereafter we shall not distinguish between the two index types, but remember instead that $\lambda_\alpha$ can  contract on the right or left with $\lambda_\beta^\dagger$, but not with $\lambda_\beta$, and that $\lambda^\dagger\lambda$ lives in a different space to $\lambda \lambda^\dagger$ etc. 

\emph{Alignment --} Suppose that 
\begin{equation}
	\label{eqn:CA}
 	[\lambda,\xi]_1 \equiv \lambda^\dagger \xi - \xi^\dagger \lambda =0\,,\quad [\lambda,\xi]_2 \equiv \lambda \xi^\dagger - \xi \lambda^\dagger=0\,,
\end{equation}
so that $\lambda^\dagger\xi$ and $\lambda\xi^\dagger$ are both Hermitian. This is necessary and sufficient to ensure that $\lambda$ and $\xi$ may be simultaneously biunitarily real diagonalized by the same two unitary matrices. I.e.
\begin{equation}
	\label{eqn:LXA}
	 \lambda = U D_\lambda V^\dagger\, \qquad \mbox{and} \qquad \xi = U D_\xi V^\dagger\,,
\end{equation}
with $D_\lambda$ and $D_\xi$ diagonal and real: we say $\lambda$ and $\xi$ are `\emph{aligned}'. This result extends to a set of $k \ge 2$ tensors $\lambda_\alpha$ that all satisfy the condition \eqref{eqn:CA} pairwise.  We include a proof in the appendix. (See also Ref. \cite{Maehara:2011sv}, which proves a more general statement.)

\emph{Spectrally Disjoint --} Suppose we further require
\begin{equation}
	\label{eqn:DAS}
	\lambda^\dagger \xi = 0\,\qquad \mbox{and} \qquad\lambda \xi^\dagger = 0\,.
\end{equation}
This condition subsumes eq.~\eqref{eqn:CA}, and so $\lambda$ and $\xi$ must be aligned. When combined with eq.~\eqref{eqn:LXA} this condition further implies $D_\lambda D_\xi = 0$, or in index notation ${d_\lambda}_i {d_\xi}_i  = 0$, for each $i$, where ${d_\lambda}_i$ and ${d_\xi}_i$ are the real diagonal elements of $D_\lambda$ and $D_\xi$. Hence under the condition~\eqref{eqn:DAS}, $\lambda$ and $\xi$ are required to be aligned and  `\emph{spectrally disjoint}', in the sense that ${d_\xi}_i = 0$ whenever ${d_\lambda}_i \not = 0$, and vice versa. The converse statement follows trivially.  Eq.~\eqref{eqn:DAS} extended pairwise to a set of $k$ tensors is therefore sufficient and necessary for them all to be aligned and spectrally disjoint. 

\emph{Rank-1 -- } A maximal set of $n$ linearly independent tensors that satisfy \eqref{eqn:DAS} pairwise, are automatically also `\emph{rank-1}', in the sense that each tensor must have a single non-zero eigenvalue. More generally, any single tensor $\lambda$ is rank-1 if and only if
\begin{align} 
	\label{eqn:R1C}
	\Tr(\lambda^\dagger\lambda)^2 - \Tr(\lambda^\dagger\lambda\lambda^\dagger\lambda) = 0\,.
\end{align}
In index notation this becomes $\sum_{i <j} |{d_\lambda}_i|^2|{d_\lambda}_j|^2=0$, and the only non-trivial solution is $|{d_\lambda}_{i_0}| > 0$ and ${d_\lambda}_{i \not=i_0} = 0$ for some $i_0 \in \{1,\ldots n\}$. Hence $\lambda$ is rank-1, and the converse argument is trivial.  A set of tensors therefore has the aligned, spectrally disjoint and rank-1 structure \eqref{eqn:ASDS} if and only if the algebraic conditions \eqref{eqn:DAS} and \eqref{eqn:R1C} are satisfied pairwise and individually on the set, respectively.

\textbf{Potential.} We now proceed to construct a potential that ensures \eqref{eqn:DAS} and \eqref{eqn:R1C} hold dynamically for a set of up-type flavons, $\lambda_{\alpha}\in\{\lambda_t,\lambda_c,\lambda_u\}$. The parities $P_\alpha: \lambda_\alpha \to -\lambda_\alpha$ restrict the form of the renormalizable potential, such that it may only involve terms containing at most two different flavons. The full potential for $k\ge2$ flavons can therefore be  constructed from a sum of single-field potentials and two-field potentials.

\emph{Single-field potential --}  The most general flavor-invariant renormalizable potential for a single flavon is
\begin{align}
	V^\alpha_{1\textrm{f}}
	& = \mu^\alpha_1 \Big| \Tr(\lambda_\alpha^\dagger \lambda^\pda_\alpha)  - r_\alpha^2 \Big|^2  \!\! \notag\\
	& \quad  + \mu^\alpha_2\! \Big[\Tr(\lambda_\alpha^\dagger \lambda^\pda_\alpha)^2- \Tr(\lambda_\alpha^\dagger \lambda^\pda_\alpha\lambda_\alpha^\dagger \lambda^\pda_\alpha)\Big]\notag\\
	& = \mu^\alpha_1 \Big| \!\sum_i \!|{d_\alpha}_i|^2 - r_\alpha^2 \Big|^2 \!\! + 2\mu^\alpha_2 \! \sum_{i <j} |{d_\alpha}_i|^2|{d_\alpha}_j|^2\,.\label{eqn:MSP}
\end{align}
Both operators are positive semi-definite. For $\mu^\alpha_{1,2} >0$, eq.~\eqref{eqn:R1C} is therefore satisfied at the minimum of this potential. The particular solution is $|{d_\alpha}_{i_0}| = r_\alpha$  for some $i_0 \in \{1,\ldots n\}$ and ${d_\alpha}_{i \not=i_0} = 0$. Hence $\lambda_\alpha$ is rank-1.

\emph{Two-field potential --} The most general CP-conserving, flavor- and parity-invariant renormalizable potential for two fields $\lambda_\alpha$ and $\lambda_\beta$ can be written as
\begin{align}
	& V_{2\textrm{f}}^{\alpha\beta} = \mu^{\alpha\beta}_3 \! \Big[ \Tr(\lambda_\alpha^\dagger\lambda^\pda_\alpha) + \Tr(\lambda_\beta^\dagger\lambda^\pda_\beta) - r_\alpha^2 - r_\beta^2\Big]^2 \label{eqn:AMS} \\
	& \quad + \sum_{\pm} \mu^{\alpha\beta}_{4,\pm} \Big|\Tr\Big[\lambda_\alpha^\dagger \lambda^\pda_\beta \pm \lambda_\beta^\dagger \lambda^\pda_\alpha\Big]\Big|^2 \notag\\
	& \quad + \sum_{i = 1,2}\mu^{\alpha\beta}_{5,i} \Tr\Big[ [\lambda_\alpha,\lambda_\beta]_{i}^\dagger [\lambda_\alpha,\lambda_\beta]^\pda_{i}\Big]\notag \\
	 &\quad + \mu^{\alpha\beta}_6 \! \Big\{\! \Tr\Big[(\lambda^\pda_\alpha\lambda_\beta^\dagger)^\dagger(\lambda^\pda_\alpha\lambda_\beta^\dagger)\Big] +\Tr\Big[(\lambda_\alpha^\dagger\lambda^\pda_\beta)^\dagger(\lambda_\alpha^\dagger\lambda^\pda_\beta)\Big] \! \Big\} \nonumber \,.
\end{align}
The operators in eq.~\eqref{eqn:AMS} are all manifestly positive semi-definite. With all the coefficients positive, the global minimum of the potential is thus $V_{\textrm{2f}}=0$. The operator corresponding to $\mu_3$ vanishes if both $\lambda_\alpha$ and $\lambda_\beta$ are in the vacua of their single field potentials~\eqref{eqn:MSP}. 
The operator corresponding to $\mu_6$ is non-zero \emph{if and only if} $\lambda_\alpha^\dagger \lambda^\pda_\beta = \lambda^\pda_\alpha \lambda_\beta^\dagger = 0$, and all remaining operators also vanish at this condition. Hence the global minimum of $V_{\textrm{1f}}$ and $V_{\textrm{2f}}$ together is located at the aligned, spectrally disjoint and rank-1 conditions~\eqref{eqn:DAS} and \eqref{eqn:R1C}. 

Extended to a set of $k$ fields, $\{\lambda_\alpha\}$, the pairwise potential 
\begin{equation}
	\label{eqn:PP}
	V_{\textrm{pp}} = \sum_{\alpha} V_{1\textrm{f}}^\alpha  + \sum_{\alpha<\beta} V_{2\textrm{f}}^{\alpha\beta}\,,
\end{equation}
with couplings all positive thus dynamically generates a set of spurions of the desired pattern~\eqref{eqn:ASDS}. The flat directions of its minimum are parametrized solely by the unitary matrices $U$ and $V$, which simultaneously rotate $\{\lambda_\alpha\}$ as in \eqref{eqn:ASDS}. Although the potential appears to contain a very large number of free parameters, the only significant parameters for the low energy physics are the radial norms $r_\alpha$, as long as all other parameters are positive.

\emph{Parity breaking effects --} Breaking of the $P_\alpha$ symmetries in the hierarchy sector can radiatively induce parity-odd operators in the potential, e.g.~$\Tr(\lambda_\alpha^\dagger \lambda^\pda_\beta)$. Since all such operators are invariant under the simultaneous rotation of the set $\{\lambda_\alpha\}$, they do not destabilize the flat directions of the vacuum. In addition, these parity-odd terms are suppressed by $(\langle s_\alpha \rangle\langle s_\beta\rangle/ \LambdaH^2)$ for every pair of parity symmetries $P_{\alpha,\beta}$ that they break, and can be further two-loop suppressed by the SM portal \eqref{eqn:SMY} (see the UV completion in Fig.~\ref{fig:GQ} for an example). For the SM quarks, the largest parity-odd contribution is then $\sim (m_cm_t/v^2)/(16\pi^2)^2 \ll m_u/m_t$, the largest hierarchy in the system.  All such terms may then be neglected.

\emph{Two-sector potential -- }  Now consider a second set of three down-type flavons $\lambda_{\hat\alpha} \in \{\lambda_b,\lambda_s,\lambda_d\}$, that are charged under flavor $U(3)_Q\times U(3)_D$. We distinguish these from the up-type flavons by their hatted index. The common $U(3)_Q$ group admits up-down cross terms
\begin{align}
	\label{eqn:CT}
	V_{\textrm{mix}}^{\alpha \hat\alpha} 
	& = \nu_1^{\alpha \hat\alpha} \Tr \Big[(\lambda^\dagger_\alpha \lambda^\pda_{\hat\alpha})^\dagger (\lambda^\dagger_\alpha \lambda^\pda_{\hat\alpha})\Big]\notag\\ 
	& \quad +\nu_2^{\alpha \hat\alpha}\Big[\Tr (\lambda_\alpha^\dagger \lambda^\pda_\alpha)+\Tr (\lambda_{\hat\alpha}^\dagger \lambda^\pda_{\hat\alpha})-r_\alpha^2 -r_{\hat\alpha}^2\Big]^2\,,
\end{align}
into the most general CP-, flavor- and parity-invariant potential, i.e. $V_{\textrm{pp}}^{\textrm{up}} + V_{\textrm{pp}}^{\textrm{down}} + V_{\textrm{mix}}$. Both operators are positive semi-definite, and we assume $\nu_{1,2}>0$. The $\nu_2$ term vanishes at the vacua of $V_{\textrm{pp}}$, but the $\nu_1$ terms cannot vanish simultaneously with the $\mu_{6}$ terms, since one cannot non-trivially satisfy $\lambda^\dagger_\alpha \lambda^\pda_\beta = \lambda^\dagger_\alpha\lambda^\pda_{\hat\alpha} = \lambda^\dagger_{\hat\alpha} \lambda^\pda_{\hat\beta} = 0$.

Since $V_{\textrm{mix}}$ respects $\lambda^\pda_\alpha \lambda_\beta^\dagger \to -\lambda^\pda_\alpha \lambda_\beta^\dagger$ for $\alpha\neq \beta$, it cannot introduce tadpoles that shift the non-trivial stationary points of $V_{\textrm{pp}}$ from the $\{\lambda^\pda_\alpha \lambda_\beta^\dagger = 0\}_{\alpha \not=\beta}$ contour. Moreover, the $\nu_1$ term has curvature $\partial^2 V_{\textrm{mix},\nu_1}/\partial \lambda^\pda_\alpha \partial \lambda_\beta^\dagger \propto \delta_{\alpha\beta}$. Provided $\nu_1$ is somewhat small compared to $\mu_{6}$ and $\mu_{4,+}$, this term cannot destabilize an existing $V_{\textrm{pp}}$ minimum.  No symmetries, however, forbid tadpoles that shift the location of the radial vacuum $\Tr(\lambda^\pda_\alpha\lambda^\dagger_\alpha)$.  Hence the total potential retains local non-trivial minima somewhere on the $\{\lambda^\pda_\alpha \lambda_\beta^\dagger = 0\}_{\alpha \not=\beta}$ contour, i.e. at the aligned, spectrally disjoint configuration.   For $\nu^{\alpha\hat\alpha}_1 >0$, the cross terms typically squeeze the location of the radial vacuum to $\Tr(\lambda^\pda_\alpha\lambda^\dagger_\alpha) = \bar{r}^2_\alpha < r^2_\alpha$. Provided $\nu_1^{\alpha\hat\alpha}$ are not too large compared to the $\mu_{1}^\alpha$ terms, the vacuum remains non-trivial, i.e. $\langle \lambda_\alpha \rangle \not= 0$. 

It remains to check that the unit rank of $\langle \lambda_\alpha \rangle$ is not spoiled. The desired configuration \eqref{eqn:ASDS} is explicitly
\begin{equation}
	\label{eqn:UDC}
	\lambda_\alpha = U^\pda_U D_\alpha V_U^\dagger\,, \qquad \lambda_{\hat\alpha} = U^\pda_D D_{\hat\alpha} V_D^\dagger\,,
\end{equation}
where we choose $D_\alpha$ to be the rank-1 diagonal matrix whose $\alpha$-th diagonal entry, $d_\alpha \not= 0$. At this configuration the cross terms become
\begin{equation}
	\label{eqn:EM}
	V_{\textrm{mix}}^{\alpha \hat\alpha} = \nu_1^{\alpha \hat\alpha} d_\alpha^2 d_{\hat\alpha}^2 \big| \mathcal{V}_{\textrm{ckm}}^{\alpha\hat\alpha}\big|^2\,,
\end{equation}
where $ \mathcal{V}_{\textrm{ckm}}^{\alpha\hat\alpha}$ is the $\alpha\hat\alpha$-th element of $\mathcal{V}_{\textrm{ckm}} \equiv U_D^\dagger U^\pda_U$, the usual unitary up-down mixing matrix. Unitarity forbids all these terms from being simultaneously zero. This term also lifts the $U_U$ and $U_D$ flat directions of the potential \eqref{eqn:PP}. That is, it determines the texture of $\mathcal{V}_{\textrm{ckm}}$. 

Perturbing the $\beta$th diagonal zero entry of $D_\alpha$ by $\epsilon$ corresponds to perturbing the rank-1 (or disjoint) configuration. From eqs.~\eqref{eqn:CT} and \eqref{eqn:UDC} this generates only an $\mathcal{O}(\epsilon^2)$ correction $\delta V_{\textrm{mix}}^{\alpha \hat\alpha} = \epsilon^2\nu_1^{\alpha \hat\alpha}d_{\hat\alpha}^2 | \mathcal{V}_{\textrm{ckm}}^{\beta \hat\alpha}|^2$. One may similarly check that in the vacuum of eq.~\eqref{eqn:EM}, $\mathcal{O}(\epsilon)$ perturbations of the alignment condition arise in $V_{\textrm{mix}}$ at $\mathcal{O}(\epsilon^2)$, in concordance with the argument above. Thus, provided $\mu_{6}$, $\mu_{4,+} \gtrsim \nu_1 > 0$, there remains a local minimum at the aligned, spectrally disjoint, rank-1 configuration for each set. In contrast, note that perturbing the non-zero element $d_\alpha \to d_\alpha + \epsilon$ leads to a $\mathcal{O}(\epsilon)$ tadpole, as above, that shifts the radial vacuum from away from $r_\alpha$.

\textbf{SM Hierarchies.} 
\emph{Quark Sector --} In general, one is free to choose the mechanism at work in the hierarchy sector. We present here an example which makes use of horizontal discrete symmetries to generate the SM quark hierarchies.

We assign an integer charge $p_\alpha$ ($p_{\hat\alpha}$) to each $\lambda_\alpha$ ($\lambda_{\hat\alpha}$) under its own individual discrete symmetry $\mathbb{Z}_{2p_\alpha}$ ($\mathbb{Z}_{2p_{\hat\alpha}}$), except for $\lambda_t$. These discrete symmetries act as the parity symmetries $P_\alpha$ on the flavons, required to secure the potential in eqs.~\eqref{eqn:PP} and \eqref{eqn:CT}.\footnote{One could have instead assigned a unique $U(1)_\alpha$ symmetry to each $\lambda_\alpha$. This generates the same potential, but with extra suppressions that reduce the $\mu_{4,\pm}$, $\mu_{5,i}$ and $\mu_6$ terms to $\mu_{6,1}\Tr[(\lambda^\pda_\alpha\lambda_\beta^\dagger)^\dagger(\lambda^\pda_\alpha\lambda_\beta^\dagger)] +\mu_{6,2}\Tr[(\lambda_\alpha^\dagger\lambda^\pda_\beta)^\dagger(\lambda_\alpha^\dagger\lambda^\pda_\beta)]$. The vacuum remains unchanged for $\mu_{6,i} >0$.} The suppression of parity-odd terms is not spoiled if only a single flavon -- $\lambda_t$ in this case -- in each set does not carry a parity. For each symmetry $\mathbb{Z}_{2p_\alpha}$ ($\mathbb{Z}_{2p_{\hat\alpha}}$) we further assign a field $\sigma_\alpha$ ($\sigma_{\hat\alpha}$), belonging to the hierarchy sector, with unit discrete charge. This produces the irrelevant operators
\begin{align}
	& H^\dagger \bar{Q}_L\! \Bigg\{ \! 
	\frac{\lambda_t}{\LambdaF} +\bigg[\!\frac{\sigma_c}{\LambdaH}\!\bigg]^{\!p_c} \!\! \frac{\lambda_c}{\LambdaF} + \bigg[\!\frac{ \sigma_u}{\LambdaH}\!\bigg]^{\!p_u} \!\! \frac{\lambda_u}{\LambdaF}\!\Bigg\} U_R\,\notag\\
	&+H \bar{Q}_L \!\Bigg\{\! \bigg[\!\frac{\sigma_b}{\LambdaH}\!\bigg]^{\!p_b} \!\! 
	\frac{\lambda_b}{\LambdaF} +\bigg[\!\frac{\sigma_s}{\LambdaH}\!\bigg]^{\!p_s} \!\! \frac{\lambda_s}{\LambdaF} + \bigg[\!\frac{ \sigma_d}{\LambdaH}\!\bigg]^{\!p_d} \!\! \frac{\lambda_d}{\LambdaF}\!\Bigg\} D_R\,.
\end{align}
There is no $\mathbb{Z}_{2p_t}$ nor $\sigma_t$, so that the top Yukawa is unsuppressed. Applying the pairwise potential \eqref{eqn:PP} and \eqref{eqn:CT} to both up- and down-type flavons, we obtain a complete set of aligned, spectrally disjoint and rank-1 spurions as in eq.~\eqref{eqn:UDC}. I.e. $D_t = \mbox{diag}\{0,0,\bar{r}_t\}$ and so on, with $\bar{r}_\alpha \lesssim \LambdaF$ the radial location of the vacuum. 

If we further assume an approximately uniform scale of breaking for all the discrete symmetries $\langle \sigma_\alpha \rangle/\LambdaH  \sim \varepsilon $  --  a natural assumption --  then an $\langle s_\alpha \rangle \sim \LambdaH \varepsilon^{p_\alpha}$ hierarchy is generated by the discrete charges $p_\alpha$ alone. For example, one could make the discrete charge choices 
\begin{equation}
p_c = 2\,,~~p_u = 5\,,~~p_b =2\,,~~p_s = 3\,,~~p_d = 5\,.
\end{equation}
For $\varepsilon \sim 0.1$, this approximately reproduces the SM quark mass hierarchies. 

For anarchic $\nu_1^{\alpha\hat\alpha}>0$, the potential \eqref{eqn:EM} ensures that the flavor mixing matrix settles to a sparse unitary matrix. One can, however, generate an approximation of the observed CKM matrix with some special choices. Suppose there exists a symmetry which requires the couplings $\nu_i^{\alpha\hat\alpha}$ and $\mu_i^{\alpha\beta}$ to be universal in $\alpha$ and $\hat\alpha$, while $r_t > r_u = r_c$ and $r_b > r_d = r_s$. One may show the potential is minimized for a mixing matrix
\begin{equation}
	\mathcal{V}_{\textrm{ckm}} = \begin{pmatrix} \phantom{-}\cos\theta & \sin\theta & 0~ \\ -\sin\theta &  \cos\theta  & 0~ \\ 0 & 0 & 1~ \end{pmatrix}\,,
\end{equation}
which has a single, arbitrarily large, mixing angle for the first two generations. Reproducing the rest of the CKM likely requires the introduction of further small perturbations, perhaps arising from irrelevant operators or interactions coupling to $U(3)_U\times U(3)_D$. We emphasize that this flavor-violating physics is independent from the dynamics of the hierarchy sector.

\emph{Lepton Sector --} A similar process may be applied to the SM leptons, for $H\bar{L}_L E_R$ and $H^\dagger \bar{L}_L N_R$ yukawas analogous to \eqref{eqn:SMY}. For $\nu_i^{\alpha\hat\alpha}$, $\mu_i^{\alpha\beta}$, and $r_\alpha$ all universal  in $\alpha$ and $\hat\alpha$, the PMNS mixing matrix may have arbitrary $\mathcal{O}(1)$ entries. (A different mechanism, however, may be responsible for the extreme overall suppression of the neutrino yukawas.)

The degeneracy of two of the neutrino masses in the case of an inverted hierarchy \cite{Gando:2013nba} can be explained if this sector has only two flavons, $\lambda$ and $\xi$, but with $\mu^\xi_2 < 0$, $2\mu_{4,+} + \mu_6 \gg  |\mu^\xi_2|/2$, and still $\mu^\lambda_2>0$. In this scenario, $\mu^\xi_2 <0$ relaxes the rank-1 condition, such that the $\xi$ spurion eigenspectrum prefers instead to be degenerate. When combined with the more energetically favored disjoint condition \eqref{eqn:DAS} enforced by $\mu_6$ and $\mu_{4,+}$, one finds
\begin{align}
	\langle \lambda \rangle/\LambdaF &\sim U\,\mbox{diag}\Big\{0,\,0,\,1\Big\}\,V^\dagger\,,\notag\\
	\langle \xi \rangle/\LambdaF &\sim U\,\mbox{diag}\Big\{1,\,1,\,0\Big\}\,V^\dagger\,.
\end{align}
One may then obtain two degenerate Dirac neutrino masses and one much lighter.

\textbf{BSM Applications.}
In the context of beyond SM (BSM) model building, it is often desirable to obtain new physics (NP) whose flavor-breaking effects are aligned with, but not proportional to, the SM yukawas. This is more general than minimal flavor violation, and can be achieved dynamically with the $V_{\textrm{pp}}$ potential.

Assume the existence of an SM spurion $\lambda_{\textrm{sm}} \sim U\,\mbox{diag}\{\delta',\delta,1\}\,V^\dagger$, with $\delta' \ll \delta \ll 1$, and a second field $\lambda_{\textrm{np}}$ whose vacuum expectation value represents a flavor-breaking NP spurion. We apply the potential \eqref{eqn:PP} for these two spurions, but treat $\lambda_{\textrm{sm}}$ as a static background field, fixed by some high scale physics. For $\lambda_{\textrm{sm}}$ and $\lambda_{\textrm{np}}$ to be aligned, it suffices that the condition \eqref{eqn:CA} is satisfied:  If $\mu_{5,i} >0$ it is energetically favorable for $\lambda_{\textrm{np}}$ to settle such that the corresponding operators vanish. This automatically results in alignment with $\lambda_{\textrm{sm}}$. 

Since $\lambda_{\textrm{sm}}$ has maximal rank, it is not possible for the two spurions to be spectrally disjoint. In the limit $\mu_1^{\textrm{np}}> |\mu_2^{\textrm{np}}|\gg\mu_{4,+},\; \mu_6$, taking all constants positive except $\mu_2^{\textrm{np}}$, the vacuum solution is either one of
\begin{align}
	\langle \lambda_{\textrm{np}} \rangle/\LambdaF & \sim U\, \mbox{diag}\Big\{1,\,0,\,0\Big\}\, V^\dagger\,,\notag\\
	\langle \lambda_{\textrm{np}} \rangle/\LambdaF & \sim U\, \mbox{diag}\Big\{1,\,1,\,1\Big\}\, V^\dagger\,,
\end{align}
corresponding to whether $\mu_2^{\textrm{np}}>0$ or $\mu_2^{\textrm{np}}<0$ respectively. These two spurions are linearly independent and aligned. As such one can span the whole space of possible aligned NP spurions by taking linear combinations of these two spurions and $\lambda_{\textrm{sm}}$.

\textbf{Conclusions.}
We have shown that the SM fermion mass and mixing angle hierarchies may have autonomous origins, such that they may arise at vastly different physical scales. This result is a consequence of a new mechanism, in which the vacuum of the general flavon field potential dynamically generates an aligned, spectrally disjoint, and rank-1 structure for the $U(3)\times U(3)$ flavor-breaking spurions. Of particular significance, this mechanism permits the physics responsible for the SM quark mass or yukawa hierarchies to operate close to the electroweak scale, without being in conflict with precision flavor constraints. It may therefore be experimentally accessible at LHC. It also may have broad applications in the construction of flavor-safe, natural BSM theories or for electroweak baryogenesis.

\textbf{Acknowledgments.} We thank Bob Cahn, Yuval Grossman, Lawrence Hall, Zoltan Ligeti, Aneesh Manohar, Yasunori Nomura, Duccio Pappadopulo and Michele Papucci for helpful conversations. The work of SK was supported by the LDRD Program of LBNL under U.S. Department of Energy Contract No. DE-AC02-05CH11231. DR is supported by the NSF under grant No.~PHY-1002399.

\textbf{Appendix.} 
In this appendix we provide a proof of the equivalence of eqs. \eqref{eqn:CA} and \eqref{eqn:LXA}.

Repeatedly applying the condition \eqref{eqn:CA} implies that
\begin{align}
	\big[ \lambda^\dagger \lambda, \xi^\dagger \xi \big] & =0\,, \notag\\
	\big[ \lambda\lambda^\dagger, \xi\xi^\dagger \big] & =0\,.\label{eqn:ACA}
\end{align}
Since $\lambda^\dagger\lambda$ and $\xi^\dagger \xi$ are manifestly Hermitian and commute, there exists a unitary matrix $V$ such that $\lambda^\dagger \lambda = V  |D_{\lambda}|^2 V^\dagger$, $\xi^\dagger \xi= V |D_{\xi}|^2 V^\dagger$, where $|D_{\lambda,\xi}|^2$ are real positive semidefinite diagonal matrices. Similarly, since $\lambda\lambda^\dagger$ and $\xi\xi^\dagger$ are manifestly Hermitian and commute, there exists a unitary matrix $U$ such that $\lambda \lambda^\dagger = U  |D_{\lambda}|^2 U^\dagger$ and $\xi \xi^\dagger = U  |D_{\xi}|^2 U^\dagger$. It immediately follows that $\lambda$ and $\xi$ have simultaneous singular value decompositions
\begin{equation}
	\lambda = U D_{\lambda} V^\dagger~,\qquad \mbox{and} \qquad \xi= U D_{\xi} V^\dagger~,
\end{equation}
where $D_{\lambda,\xi}$ are diagonal matrices. It remains to show they are real.

We may always choose $U$ and $V$ such that $D_\lambda$ has real entries, i.e. ${d_\lambda}_i \in \mathbb{R}$. Reapplying the commutation condition \eqref{eqn:CA} implies that, in index notation 
\begin{equation}
	 {d_\lambda}_i \big({d^{\phantom{*}}_\xi}_i -  {d_\xi^*}_i\big) = 0\,,
\end{equation}
for each $i$. Hence, whenever ${d_\lambda}_i >0$, ${d_\xi}_i$ is real too. Whenever ${d_\lambda}_i = 0$, we may always absorb the phase of ${d_\xi}_i$ into $U$ by means of a unitary left-transformation $D_\alpha \mapsto X D_\alpha$ that leaves $D_\lambda$ invariant. So there exists $U$ and $V$ that simultaneously real diagonalize $\lambda$ and $\xi$. The converse argument -- that \eqref{eqn:LXA} implies \eqref{eqn:CA} -- is trivial.

This argument extends by strong induction to multiple tensors. Suppose a set $k$ tensors $\{\lambda_\alpha\}$ satisfy eqs.~\eqref{eqn:CA} pairwise. Then it follows from eqs.~\eqref{eqn:ACA} that they are all simultaneously biunitarily diagonalizable by unitary matrices $U$ and $V$. Now further suppose that the first $k-1$ are diagonalized by $U$ and $V$ into real diagonal matrices, i.e. $D_{\alpha < k}$ are real. Then for the $k$th tensor, $D_k$, ${d_\alpha}_i \big({d^{\phantom{*}}_k}_i -  {d^{*}_k}_i\big) = 0$, for all $\alpha = 1,\ldots,k-1$. Whenever any of the ${d_\alpha}_i$ are non-zero, then ${d_k}_i$ must be real too. If ${d_{\alpha}}_i$ are zero for a particular $i$ and for all $\alpha < k$, then one may absorb the phase of ${d_k}_i$  into $U$ by means of a unitary left-transformation $D_\alpha \mapsto X D_\alpha$ that leaves $D_{\alpha <k}$ invariant. Hence there exists $U$ and $V$ such that $D_{k}$ is real too.

%\bibliography{FlavorHierarchies}


\end{document}